\documentclass[manuscript,screen]{acmart}
\AtBeginDocument{%
  }

\setcopyright{acmlicensed}
\copyrightyear{2026}
\acmYear{2026}
\acmDOI{XXXXXXX.XXXXXXX}
\acmConference[Conference acronym 'XX]{Make sure to enter the correct
  conference title from your rights confirmation emai}{June 03--05,
  2018}{Woodstock, NY}
\acmISBN{978-1-4503-XXXX-X/18/06}

\usepackage{xspace}
\newcommand\fwo{\texttt{FMware Runtime}\xspace}
\usepackage{soul}
\setuldepth{Berlin}
\usepackage{hyperref}
\hypersetup{colorlinks=true,breaklinks=true}
\usepackage{multirow}
\usepackage{makecell}
\usepackage{balance}
\usepackage{enumitem}
\usepackage{pifont}
\usepackage{algorithm}
\usepackage{algpseudocode}

\newboolean{showcomments}
\setboolean{showcomments}{true}

\ifthenelse{\boolean{showcomments}}
{\newcommand{\nbc}[3]{
 {\colorbox{#3}{\bfseries\sffamily\scriptsize\textcolor{white}{#1}}}
 {\textcolor{#3}{\sf\small$\blacktriangleright$\textit{#2}$\blacktriangleleft$}}
 }
}
{\newcommand{\nbc}[3]{}
 }

\begin{document}

%%
%% The "title" command has an optional parameter,
%% allowing the author to define a "short title" to be used in page headers.
\title{Software Performance Engineering for Foundation Model-Powered Software}

%%
%% The "author" command and its associated commands are used to define
%% the authors and their affiliations.
%% Of note is the shared affiliation of the first two authors, and the
%% "authornote" and "authornotemark" commands
%% used to denote shared contribution to the research.
\author{Haoxiang Zhang}
\affiliation{%
  \institution{Centre for Software Excellence at Huawei}
  \country{Canada}
}
\email{cse@huawei.com}

\author{Shi Chang}
\affiliation{%
  \institution{Western University}
  \country{Canada}}

\author{Arthur Leung}
\affiliation{%
  \institution{Centre for Software Excellence at Huawei}
  \country{Canada}
}

\author{Kishanthan Thangarajah}
\affiliation{%
  \institution{Centre for Software Excellence at Huawei}
  \country{Canada}
}

\author{Boyuan Chen}
\affiliation{%
  \institution{Centre for Software Excellence at Huawei}
  \country{Canada}
}

\author{Hanan Lutfiyya}
\affiliation{%
  \institution{Western University}
  \country{Canada}
}

\author{Ahmed E. Hassan}
\affiliation{%
  \institution{Queen's University}
  \country{Canada}
}

%%
%% By default, the full list of authors will be used in the page
%% headers. Often, this list is too long, and will overlap
%% other information printed in the page headers. This command allows
%% the author to define a more concise list
%% of authors' names for this purpose.
\renewcommand{\shortauthors}{Zhang et al.}

%%
%% The abstract is a short summary of the work to be presented in the
%% article.
\begin{abstract}
The rise of Foundation Models (FMs) like Large Language Models (LLMs) is revolutionizing software development. Despite the impressive prototypes, transforming FMware into production-ready products demands complex engineering across various domains. A critical but overlooked aspect is performance engineering, which aims at ensuring FMware meets performance goals such as throughput and latency to avoid user dissatisfaction and financial loss. Often, performance considerations are an afterthought, leading to costly optimization efforts post-deployment. FMware's high computational resource demands highlight the need for efficient hardware use. Continuous performance engineering is essential to prevent degradation. This paper highlights the significance of Software Performance Engineering (SPE) in FMware, identifying four key challenges: cognitive architecture design (i.e., the structural design that defines how AI components interact, reason, and interface with classical software components), communication protocols, tuning and optimization, and deployment. These challenges are based on literature surveys and experiences from developing an in-house FMware system. We discuss problems, current practices, and innovative paths for the software engineering community.
\end{abstract}

%%
%% The code below is generated by the tool at http://dl.acm.org/ccs.cfm.
%% Please copy and paste the code instead of the example below.
%%


%%
%% Keywords. The author(s) should pick words that accurately describe
%% the work being presented. Separate the keywords with commas.

\received{13 February 2025}
\received[revised]{11 November 2025}
\received[accepted]{18 January 2026}

%%
%% This command processes the author and affiliation and title
%% information and builds the first part of the formatted document.
\maketitle

\section{Introduction}
\label{sec:intro}
The rapid emergence of Foundation Models (FMs), particularly Large Language Models (LLMs), is reshaping software development, with market value expected to reach \$36.1 billion by 2030~\cite{marketsandmarkets2024}.
FMs empower the creation of intelligent software, which we refer to as FMware (i.e., software built using FMs)~\cite{hassan2024rethinking}, where applications rely on one or more building blocks that are FMs.

Demos built with FMware reveal its profound impact across domains such as Software Engineering (e.g., code generation), Knowledge Retrieval (RAG systems), and Autonomous Task Execution, enabling a new class of versatile and intelligent applications~\cite{zhang2023automl, yang2024autommlab}. However, developing FMware from prototypes into production-ready products is a complex engineering process, requiring collaborations across AI, systems, software engineering, and hardware domains throughout the lifetime of such software~\cite{oreilly2024llmtactical,guo2024large}. 

Performance engineering, one of the key aspects in such an engineering process, has not been thoroughly discussed. That is, how to proactively ensure that the developed FMware meets the pre-defined performance goals, e.g., throughput or latency. These goals are sometimes also referred to as Service Level Agreements (SLAs) or Service Level Objectives (SLOs). Failing to meet these goals while developing FMware will result in unsatisfactory user experiences or lead to substantial cost.

However, in practice, \textbf{performance concerns are often considered afterthoughts} during the lifecycle of FMware, leading to reactive and costly optimisation efforts once SLAs are not met~\cite{griggs2024melangecostefficientlarge,10.5555/3691938.3691946}. Moreover, the intensive computational resources required to serve FMware can make deployment prohibitively expensive. Efforts to improve overall hardware utilization and avoid wastage of scarce computing resources (e.g., idle GPUs) have been highlighted as a critical need in recent work~\cite{10.5555/3691938.3691946}. Finally, since FMware is live software that evolves continuously in production, continuous performance tuning practices and monitoring are necessary to prevent performance degradation over time~\cite{oreilly2024llmtactical2}. To summarize, Software Performance Engineering (SPE) practices are crucial for bringing FMware from prototype to production. Although awareness of performance-oriented FMware production is growing~\cite{oreilly2024llmtactical,guo2024large}, systematic studies focusing specifically on SPE for FMware (SPE4FMware) remain limited.

In this paper, we present a comprehensive analysis of SPE challenges in FMware development, deriving from
% %%%
% HZ: the line below is shortened here, but extended in Section 3.1
% %%%
% four authoritative sources: (i) an extensive survey of both academic and grey literature, (ii) in-depth discussions with industrial stakeholders and active academicians during \textit{SEMLA 2023 \& 2024}~\cite{semla2024}, \textit{FM+SE Vision 2030}~\cite{vision2023}, \textit{FM+SE Summit 2024}~\cite{summit2024}, and \textit{SE 2030 workshop - FSE 2024}~\cite{seworkshop2024} events, (iii) close collaboration with our customers and our internal FMware application development teams to understand their pain points with performance issues, and (iv) our hands-on experience designing and implementing an in-house FMware serving system (\fwo).
various sources by a thorough approach and aiming to reduce any biased opinion.
We identify four key SPE challenges that span across the lifecycle of FMware development: the design of cognitive architectures\footnote{In this work, we use the term cognitive architecture to refer to the structural design that defines how different AI components interact and reason together to achieve desired outcomes. Establishing a well-structured cognitive architecture is foundational for developing FMware, as it directly affects performance, scalability, and system integration.}, defining communication protocols, tuning and optimization approaches, and deployment options.
For each challenge, we describe its aspects in detail, discuss state-of-practices, and share our vision of innovation paths that call for contributions from the software engineering research community.

This paper is organized as follows: Section~\ref{sec:background} outlines the background of our study. Section~\ref{sec:challenges} delves into the SPE challenges that are associated with FMware. Section~\ref{sec:runtime} describes the vision of our serving system. Finally, Section~\ref{sec:conclusion} summarizes our insights and conclusions.

\section{Background}
\label{sec:background}
In this section, we first review Software Performance Engineering (SPE) research for traditional software (Section~\ref{sec:background:spe}). Then we explain the inference process of FM (Section~\ref{sec:background:inference}). We also provide an overview of SPE for FM (Section~\ref{sec:background:spe4fm}). At last, we present the background of FMware (Section~\ref{sec:background:fmwaredev}).

\subsection{Software Performance Engineering (SPE)}
\label{sec:background:spe}

Software Performance Engineering (SPE) was introduced in the early 1980s for systematically addressing performance concerns early and continuously throughout the software development lifecycle~\cite{Smith1981_SPE}. Formally, SPE is defined as a quantitative, model-driven approach to predicting, evaluating, and ensuring that software systems meet performance objectives~\cite{SmithWilliams2002_PerformanceSolutions}. Over the decades, SPE has evolved to support object-oriented, distributed, and cloud-native systems, demonstrating its adaptability to emerging software paradigms while maintaining its focus on early prediction and lifecycle-wide performance assurance~\cite{SmithWilliams2002_PerformanceSolutions}. SPE is formalized to involve modeling and analyzing software systems to understand performance characteristics and uncover optimization opportunities~\cite{woodside2007future}. SPE encompasses various engineering practices aimed at meeting performance requirements such as latency, throughput, and resource utilization. As software complexity escalates, addressing performance issues as afterthoughts becomes increasingly challenging and costly~\cite{smith2002performance}. Hence, proactively applying SPE practices and embedding them throughout the development lifecycle is advantageous.

Traditionally, most software components are considered deterministic, allowing developers to recreate issues when diagnosing performance degradation. However, as software systems evolve with increasingly complex interactions, non-deterministic behaviours have emerged~\cite{jain1990art}. For example, in real-time embedded systems, interrupt-driven interactions with environments introduce unpredictability, as interrupts occur randomly and are handled by priority levels, making overall system behaviour non-deterministic. This non-deterministic behaviour complicates performance engineering, making it difficult to reproduce performance issues that occur in production. This challenge is amplified with FMs due to the probability-based token sampling process in FM inference. Consequently, the rise of FMware necessitates rethinking methods to accurately predict and optimize system performance.

\subsection{Foundation Models (FMs) \& Inference Process}
\label{sec:background:inference}

FMs have transformed software by providing unparalleled abilities in comprehending and generating diverse data types. Trained on extensive unlabeled datasets, these models exhibit extraordinary versatility across various tasks, ranging from natural language processing to image generation~\cite{bommasani2021opportunities}. Particularly notable are Large Language Models (LLMs), recognized for their sophisticated capabilities in text generation, language comprehension, and multilingual processing. %~\cite{brown2020language}
A notable FM architecture is the Generative Pre-trained Transformer (GPT), which employs a decoder-only model architecture, excelling in language understanding and generation. As the number of parameters scales into the billions, the model's capabilities expand to handle general tasks due to emergent behaviours~\cite{kaplan2020scaling}. A prime example is the renowned GPT-4 model by OpenAI~\cite{openai_gpt4}. 

The inference process of an FM comprises two phases: prefill and decode~\cite{agrawal2023sarathi}. In the prefill phase, the user-provided input, or prompt, is fed into the model, initiating the first forward pass to generate the initial output token. This phase is computation-intensive, involving substantial parallel matrix multiplication operations.
During the decode phase, models sequentially generate tokens iteratively, where each new token is created based on all previously generated tokens. This token-by-token generation process requires storing the previously computed tokens' keys and values (known as KV cache)~\cite{kwon2023efficient} to speed up inference by avoiding redundant computations. The decode phase is memory-bound and cannot be fully parallelized due to data dependency and the sequential nature of token generation.

\subsection{SPE for FM Inference}
\label{sec:background:spe4fm}

The FM inference process exhibits two notable characteristics that impact its performance~\cite{sun2024llumnix}. First, the queries sent to FM show a diverse length range due to workload heterogeneity. The same task can be articulated through concise instructions or elaborate descriptions, leading to variations in both first token generation latency and KV cache memory consumption. Second, the generated tokens from FM show a diverse length range due to execution unpredictability, resulting in inference completion latency ranging from seconds to minutes and memory consumption varying from megabytes to gigabytes. These characteristics significantly impact how different performance requirements can be satisfied in practice. There are commonly three types of inference tasks: long input and short output (e.g., summarizing an essay), long input and long output (e.g., editing an essay), and short input and long output (e.g., generating an essay). Some tasks mainly demand low latency for the first output token, as subsequent token generation only needs to match human reading speed. In contrast, other tasks require minimal overall latency. These different requirements highlight the importance of performance engineering based on specific use cases.

As FMs continue to expand in size and capability, following the scaling laws~\cite{kaplan2020scaling,hoffmann2022training}, optimizing their underlying architecture is essential not only during model training~\cite{10.1145/3458817.3476209} but also during model inference, as performance optimization is critical for efficient FM deployment. Techniques such as model compression, quantization, and efficient hardware utilization are employed to balance performance with computational demands~\cite{zhou2024survey,wang2024model}. A thorough understanding of the FM inference process is critical for advancing SPE for FMware, as FMs serve as the fundamental components. While numerous surveys exist on inference optimization of FMs~\cite{zhou2024survey,yuan2024llm,wang2024model,xu2024survey,liu2024understanding,stojkovic2024towards,li2024personal,bai2024beyond},
%~\cite{li2024llm,zhou2024survey,yuan2024llm,qu2024mobile,wang2024model,xu2024survey,liu2024understanding,stojkovic2024towards,li2024personal,friha2024llm,bai2024beyond,chavan2024faster,zhou2024large}, 
our paper focuses specifically on the application-level (i.e., FMware) SPE challenges from the perspective of application developers rather than AI engineers -- in turn complementing existing efforts for model-level inference optimization.

\subsection{FM-Powered Software (FMware)}
\label{sec:background:fmwaredev}

FMs have become pivotal in AI-driven software applications, revolutionizing software engineering and serving as the backbone for a new category of software known as FM-powered software, or FMware~\cite{hassan2024rethinking}.

FMware can be classified into two categories: Promptware and Agentware. For a visual depiction of this FMware classification, readers may refer to Figure 1 in~\cite{hassan2024rethinking} that illustrates the Promptware and Agentware taxonomy. Promptware refers to a generation of software developed primarily through natural language prompts that directly utilize FMs~\cite{hassan2024rethinking}. It involves the direct use of FMs via one or many prompts. A notable example is Retrieval-Augmented Generation (RAG)-based software, which enhances output quality by combining FMs with external knowledge sources and documents to avoid issues like hallucination. %~\cite{lewis2020retrieval}
Promptware varies in complexity, from a single FM invocation in a question-answering session to a meticulously designed chain of invocations represented as Directed Acyclic Graphs (DAG) or workflows~\cite{NEURIPS2022_cot,hassan2024rethinking}. Each node in these workflows represents an individual task, and the edges represent sequential, parallel, or recursive interactions. Tasks can take the form of regular code scripts, traditional ML/DNN model invocations, or FM invocations. This approach enables the creation of compound AI systems capable of handling complex tasks through a series of well-defined steps~\cite{compound-ai-blog,hassan2024rethinking}. Agentware, on the other hand, represents an autonomous and dynamic form of FMware. In Agentware, AI agents powered by FMs can proactively interact with their environment, utilize tools, retain memories, communicate with other AI agents, and autonomously self-explore and improve themselves. While these agents can operate within an explicit workflow similar to Promptware, their true strength lies in autonomy, where researchers expect the agents to reason and develop plans with minimal human intervention. This behaviour emerges during runtime based on interactions and can only be observed through input/output trace data. Although autonomous AI agents remain an active area of research, with recent surveys providing a systematic review of their construction, diverse applications, and evaluation strategies~\cite{Wang2024_SurveyLLMAgents}, they are still in the early stages of development and require further exploration~\cite{liu2023bolaa,bouzenia2024repairagent}. With the characteristics of Promptware and Agentware in mind, we will explore the SPE challenges of FMware in the next section.

\section{Software Performance Engineering Challenges for FMware}
\label{sec:challenges}
\begin{figure}[htbp]
    \centering
    \includegraphics[width=\linewidth]{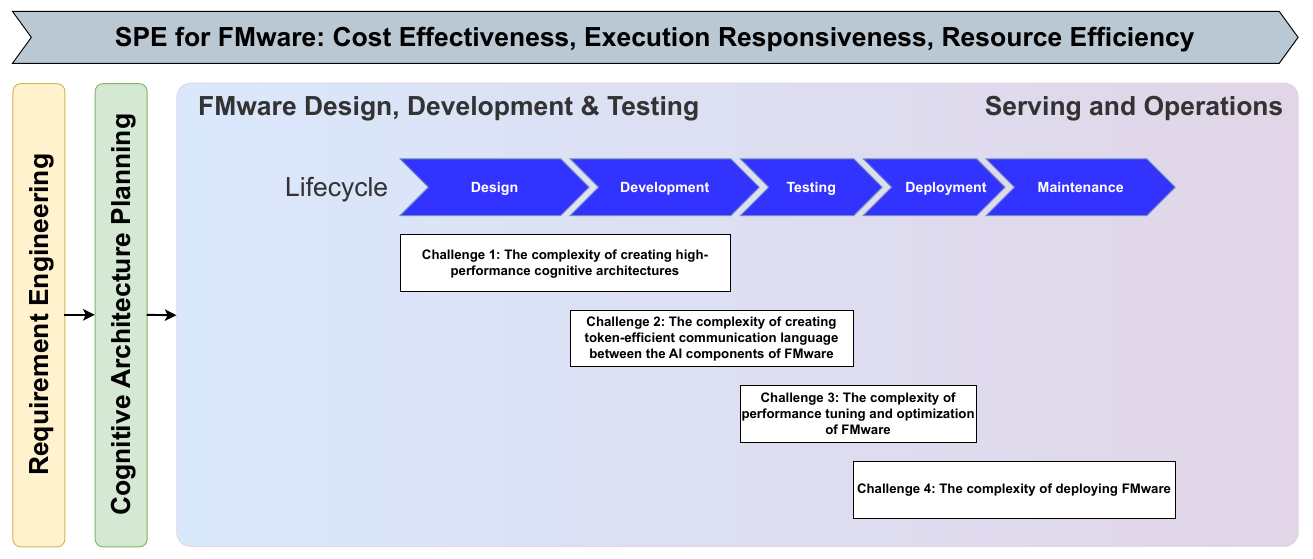}
    \caption{The illustration of software performance engineering challenges for FMware throughout the development lifecycle}
    \label{fig:challenges_map}
\end{figure}

In this section, we describe four challenges in SPE4FMware. First, we describe our methodology for deriving such challenges. Then we present the challenges in details. In particular, the following four challenges are discussed: (1) How to create a high-performance cognitive architecture for FMware (Section~\ref{challenge:architecture})? (2) How to develop a token-efficient communication language among the AI components of an FMware (Section~\ref{challenge:communication})? (3) How to continuously conduct performance tuning and optimization of FMware (Section~\ref{challenge:FM_tuning})? and (4) How to decide the deployment options for FMware (Section~\ref{challenge:deployment})? Fig.~\ref{fig:challenges_map} illustrates our four identified SPE challenges and their placement across the FMware development lifecycle. While a comprehensive description of the FMware development lifecycle is detailed in a prior work~\cite{hassan2024rethinking}, we utilize this figure to visually anchor each challenge to the specific phases where it primarily manifests. For instance, Challenges 1 and 2 are concentrated in the design, development, and testing phases, while Challenges 3 and 4 are dominant in the testing, deployment, and maintenance phases. For each challenge, we describe the characteristics unique to FMware and introduce a detailed breakdown of the challenge into several dimensions. For each dimension, we present the state of the practices that attempt to tackle the challenge and then discuss the innovation path for future research directions. 

\subsection{\textbf{Methodology}}

% %%%
% HZ: extend this subsection based on the following original line
% %%%
% four authoritative sources: (i) an extensive survey of both academic and grey literature, (ii) in-depth discussions with industrial stakeholders and active academicians during \textit{SEMLA 2023 \& 2024}~\cite{semla2024}, \textit{FM+SE Vision 2030}~\cite{vision2023}, \textit{FM+SE Summit 2024}~\cite{summit2024}, and \textit{SE 2030 workshop - FSE 2024}~\cite{seworkshop2024} events, (iii) close collaboration with our customers and our internal FMware application development teams to understand their pain points with performance issues, and (iv) our hands-on experience designing and implementing an in-house FMware serving system (\fwo).
We apply a systematic approach to derive SPE challenges for FMwaredevelopment. Our methodology for identifying these challenges follows a multifaceted, empirically grounded approach:

\begin{itemize}

    \item \textbf{Literature Review.} We conducted a systematic literature review with a search strategy encompassed both peer-reviewed academic literature and carefully selected industry sources to capture both theoretical foundations and practical insights in this rapidly evolving field. We conducted a structured search of academic paper databases (including IEEE Xplore, ACM Digital Library, and Google Scholar) using primary keywords (i.e., ``foundation model'', ``large language model'') with performance-related concepts (e.g., ``performance engineering'', ``software performance engineering'', ``inference optimization'', ``latency'', ``cost'', ``service level agreement'', ``service level objective'', etc.). To capture recent developments in this fast moving field, we supplemented with grey literature from selected arXiv preprints, industry technical reports, white papers, and engineering blogs from leading AI organizations. This mixed approach allowed us to search for foundational academic knowledge with emerging industry practices. The search was limited to the period 2022 to 2024 to ensure relevance in this fast-moving field. We applied a snowballing technique to ensure coverage and excluded papers focused purely on FM-level performance optimization that did not address FMware performance challenges. When citing non-peer-reviewed sources, we carefully qualified their scope and applicability of findings.

    \item \textbf{Community Engagement.} We actively participated in and gathered insights from in-depth discussions during premier software engineering venues, including \textit{SEMLA 2023 \& 2024}~\cite{semla2024}, \textit{FM+SE Vision 2030}~\cite{vision2023}, \textit{FM+SE Summit 2024}~\cite{summit2024}, and \textit{SE 2030 workshop - FSE 2024}~\cite{seworkshop2024} events. At these events, we conducted semi-structured interviews with industrial stakeholders and leading academicians, documenting their perspectives on performance challenges. These semi-structured interviews followed a predefined protocol centered on: 1) the most frequent performance bottlenecks, and 2) the biggest gaps in current performance engineering practices. All insights were systematically documented and reviewed by the authors to identify recurring themes and common pain points across different stakeholder groups.

    \item \textbf{Industrial Communication.} We analyzed real-world pain points related to performance through:

    \begin{itemize}
        \item Structured interviews with our customers (n=10) in the cloud application domain;
        \item Regular feedback \& debugging sessions with our internal FMware development teams, where we systematically logged all identified performance issues into a shared knowledge base;
        \item Systematic documentation of performance-related issues and incidents in a production environment.
    \end{itemize}

    \item \textbf{Industrial Experience.} We validated our findings through our hands-on experience in designing and implementing an in-house FMware serving system (\fwo). This provided practical validation of the identified challenges and their implications.

\end{itemize}

We actively designed our methodology to ensure the generalizability of the identified challenges beyond the specific context of our industrial experience. This was achieved through two critical mechanisms: (1) Data Source Selection: We synthesized insights from four independent streams—literature review, community engagement, industrial communication, and industrial experience—to identify common, foundational principles rather than project-specific performance issues. (2) Focus on Foundational SPE Gaps: The four challenges were ultimately framed as gaps in the SPE practices for the FMware development lifecycle, which are universal issues. While our internal experience provides essential empirical validation, the core challenges were validated and generalized by the common themes identified across the broader academic and industrial sources.

Following data collection, we employed a triangulation and consensus-building process to finalize the challenges. First, themes derived from the literature review were mapped against the pain points identified through community engagement and industrial communication. We then prioritized the identified challenges based on two criteria: (1) Frequency (how often a challenge appeared across all four data sources) and (2) Impact (the severity of the performance degradation as reported, when available). Challenges that were frequent and had high impact formed the basis of the four fundamental SPE challenges presented in this paper, which span the FMware development lifecycle. Our final findings were iteratively refined through reviews and validated among all the authors to ensure comprehensiveness and accuracy.

\begingroup

\subsection{\textbf{Challenge 1: The complexity of creating high-performance cognitive architectures}}
\label{challenge:architecture}

\noindent The first step in developing FMware is to create an appropriate cognitive architecture. A cognitive architecture defines how different AI components interact and reason together to achieve desired outcomes. Cognitive architectures for FMware range from simple single-FM pipelines to multi-FM orchestration such as RAG pipelines, tool-augmented chains, and workflow- or DAG-based prompt graphs. More advanced forms include autonomous agent architectures, where FM modules plan, invoke tools, and collaborate dynamically to execute multi-step reasoning and actions. The FMware cognitive architecture complements the classical software architecture, detailing how AI reasoning and results are delivered through traditional software components like regular and/or vector databases. Choices made at the cognitive architecture level can significantly impact FMware performance, either directly or through their influence on the classical software architecture. Below are some critical considerations:

\noindent{\bf Picking more powerful FMs within a simple cognitive architecture versus simpler FMs within a more complex cognitive architecture:} FMware designers face a unique performance dilemma -- they must choose between a simple cognitive architecture with fewer, larger, and more capable FMs (incurring high inference costs per request) versus a more complex cognitive architecture that combines multiple FMs (lowering inference costs per request, but involving many more inferences). We highlight that the choice of the cognitive architecture depends on the specific use case; however, generally speaking the composition of multiple FMs introduces latency and performance challenges that go beyond those encountered in traditional software.

The inference costs of FMs vary significantly, with some FM inferences being 10 times more expensive than others~\cite{llm_price_compass}. Additionally, the cost of each token generated from a single prompt is not constant. The first token incurs a much higher cost than subsequent tokens due to the need for a KV cache fill, while following tokens reuse this cache to respond faster~\cite{agrawal2023sarathi,yuan2024llminferenceunveiledsurvey}. These cost dynamics are further complicated by the introduction of the new OpenAI o1 FM, which requires more reasoning time before responding, dramatically increasing the first token's cost ~\cite{openai-o1}.

Cognitive architecture choices range from leveraging a single FM for basic interactions to complex architectures proposed in multi-agent systems ~\cite{sumers2024cognitivearchitectureslanguageagents,wu2023autogenenablingnextgenllm}. Studies and our experiences indicate that smaller FMs within a more complex cognitive architecture can achieve similar, if not better, improvements in FMware quality ~\cite{oreilly2024llmtactical,hassan2024rethinking}. However, increasing cognitive architecture complexity may result in higher latency for end-users (e.g., agents powered by weaker FMs debating each other versus a single prompt to a larger FM ~\cite{chan2023chatevalbetterllmbasedevaluators}).

Chen et al.~\cite{chen2024design} demonstrated a balanced approach to FM algorithm design, considering both error reduction and cost minimization metrics. They tuned the parallel decomposition granularity as a hyperparameter, systematically balancing competing error and performance objectives.

While complex cognitive architectures often aim to improve FMware accuracy, this may lead to suboptimal performance. Future research should explore techniques to help architects balance complex cognitive architectures with performance and cost considerations, mitigating performance overheads systematically.

\noindent{\bf Pipelining the execution of cognitive code as it is being generated versus waiting for the full generation and verification of such code:}
FMware often generates a significant portion of their source code on the fly, either by prompting an FM or through interactions with one or more AI agents. For instance, an FM might be queried to define the necessary steps (i.e., create a plan), which are then executed using FM-powered components or traditional software components. Developers can either wait for the entire set of auto-generated instructions to be completed and verified before executing them ~\cite{zhou2024isr,wang2023plan}, or start pipelining the execution, risking the need to undo steps if the overall plan is later found to be inappropriate ~\cite{chen2024toward}.

Pipelining cognitive architecture in FMware, whose code is generated on the fly, shows unique characteristics compared to classic software (Codeware).
While waiting for complete plan generation and verification ensures correctness, it introduces substantial delays (aka user-observed latencies), as post-planning execution starts only when the entire plan is generated. Pipelining execution offers better responsiveness but risks costly and complex rollbacks.

Currently, advanced mechanisms for integrating pipelining and rollbacks are implemented on a case-by-case basis without framework support, making it difficult for architects to systematically reason about such crucial and complex FMware design choices.

\noindent{\bf The addition of semantic caching throughout the cognitive architecture:} A semantic cache provides a mechanism to use prior user prompts and FM completions to address similar user prompts by using vector similarity search~\cite{Microsoft2025_SemanticCache}. Semantic caching minimizes FM or AI component inference calls by identifying similar requests or those likely to generate previously produced content. These caches are vital in optimizing the performance of FMware by reducing redundant processing and lowering latency. For example, Regmi et al.~\cite{Regmi2024_GPTSemanticCache} demonstrate that caching query embeddings in a GPT Semantic Cache can reduce LLM API calls by up to 68.8\% and significantly lower inference latency. Similarly, Gill et al.~\cite{Gill2024_PrivacyAwareSemanticCache} shows how retrieving past responses for semantically similar queries avoids repeated model invocation and reduces the latency and costs. However, designing caching mechanisms for FMware components remains ad-hoc, lacking best practices or techniques to help architects assess the ROI of adding such caches.

Typically, semantic caches utilize FMs to determine request similarity, rather than relying solely on basic text similarity metrics. This sophisticated approach enables more accurate identification of repeated or similar queries, ensuring that only necessary computations are performed. Despite their potential, the implementation of semantic caching is still in its infancy, with a need for standardized methods and frameworks to guide their development and integration into FMware.

Moreover, the effectiveness of semantic caching depends on the architecture's ability to efficiently store and retrieve cached results. This introduces challenges related to memory management and data retrieval speed, which must be addressed to realize the full benefits of semantic caching. Future research should focus on developing robust frameworks and best practices for semantic caching, ensuring that FMware can leverage these techniques to enhance performance and reduce computational overhead.

\subsection{\textbf{Challenge 2: The complexity of creating token-efficient communication language between the AI components of FMware}}
\label{challenge:communication}

\noindent Traditional software systems assume consistent communication costs between components, typically achieved through function calls or message passing via Remote Procedure Calls (RPC). For example, in a banking system, a function call might calculate interest on a savings account, taking the account balance and interest rate as inputs and returning the calculated amount. This process incurs minimal overhead due to deterministic encoding defined by the RPC interface.

However, communicating with an FM requires using natural language, which is inherently more complex. Instead of a simple function call, we must instruct the FM in natural language, e.g., ``Calculate the simple yearly interest for \$200 at an interest rate of 3.5\%.'' This approach is more verbose and inefficient, with variability in verbosity across different languages.

Parsing natural language inputs is resource-intensive compared to interpreting function calls, their parameters and return values, requiring sophisticated parsing and processing that incurs higher computational costs and latencies. Just as traditional systems use simple wire protocols for interactions, AI components need optimized communication protocols to manage their complex cognitive interactions effectively. These protocols significantly impact FMware performance.

In summary, the shift from function calls to natural language communication introduces complexity and cost, necessitating the development of specialized protocols for efficient interaction management. Below, we discuss four dimensions of this challenge in detail.

\noindent\textbf{Deciding the communication language:} 
Different natural languages require varying amounts of tokens to express the same information semantically (language efficiency and density). This disparity in word-to-token ratios across languages can significantly impact meeting performance requirements~\cite{petrov2024language}. For instance, Hindi requires eight times as many tokens as English to convey the same information~\cite{reddit2024}. This discrepancy results in longer processing times and varying performance based on the communication language used across the AI components of an FMware. API-based hosted models suffer from increased costs and longer response times with more tokens, while self-hosted models allow for language-specific fine-tuning to mitigate performance impacts.

Prior studies have sought to address the impacts of the varying word-to-token ratios. Nag et al.~\cite{nag2024cost} found that low-resource languages (LRLs) cost more than high-resource languages (HRLs) due to producing more tokens for the same content. They proposed using translation to reduce the token count processed by LRLs. However, adding translation as an intermediate step introduces drawbacks, such as increased processing time, which can affect FMware's ability to meet SLA requirements. In a prior multilingual FMware project~\cite{hassan2024rethinking}, we translated requests to English, used English for internal cognitive communication, then translated responses back. This approach improved performance despite the additional translation costs and aided developers who were not fluent in all supported languages in debugging the FMware.

Further research is needed to design multilingual applications that maintain consistent end-to-end SLAs despite token count disparities. Possible approaches include assigning powerful GPUs for LRLs to speed up processing, adopting Nag et al.'s~\cite{nag2024cost} translation step, and exploring prompt-compression techniques~\cite{jiang2023llmlingua} to reduce token counts while retaining essential information. Fine-tuning FMs can also help them better understand and process the unique characteristics of specific domains, mitigating performance impacts due to token disparities.

\noindent\textbf{Defining the communication format:} Once the communication language is decided, defining the communication format becomes crucial. JSON is a popular format, fine-tuned by many FMs for its structured, readable, and easily parsable nature~\cite{liu2024we}. However, JSON often uses more tokens than necessary to convey simple information. Alternatively, a more compact format like YAML, which is less verbose, may use fewer tokens for the same message. Using a less verbose format can make the process more efficient by reducing the time needed to process prompts and generate responses. However, format selection requires careful consideration due to FM biases towards specific output formats. Long et al.~\cite{long2024llms} found that most FMs generate correctly formatted JSON responses more reliably than YAML, likely because JSON is more prevalent in model training data. LinkedIn's shift from JSON to YAML for optimizing communication format also highlights these considerations~\cite{bottaro2024}.

Similar to human languages, grammar complexity affects performance. For instance, using verbose grammar with complex wording can negatively impact performance. Tam et al.~\cite{tam2024let} found YAML to be a more cost-effective format for models like GPT-3.5-Turbo compared to JSON, with both text and YAML formats showing lower token generation costs than JSON.

Existing research attempts to leverage less verbose formats for higher performance. Bottaro and Ramgopal~\cite{bottaro2024} noted that after switching to YAML for its brevity, FMs produced invalid output formats 10\% of the time. Some studies explored the implications of enforcing constraints on output structure. For example, Kellner et al.~\cite{beurer2024guiding} observed performance degradation with structural constraints but proposed speculative decoding to minimize overhead and speed up generation.

Chen et al.~\cite{chen2024beyond} found that using structured formats like JSON objects, tables, and markdown enhances clarity, accuracy, and reasoning efficiency in FMs, simplifying cognitive architecture and reducing error-handling needs. They proposed AutoForm, an automatic method to select and use the most suitable communication format for a task. Kurt~\cite{willkurt2024} suggested using finite-state machines and regular expressions to enforce structural constraints, improving structured output generation. Nonetheless, further work is needed to reduce invalid outputs across different schemas.

\noindent\textbf{Correcting communication messages:} Once the communication language and format structure have been defined, it is essential to ensure that the communication follows these rules. For example, if you are communicating in English and using JSON format, your messages need to be structured correctly to ensure FMs can parse and respond correctly. But if the output format is invalid or partially correct, then the downstream components of FMware will not work as expected as they may fail to understand the input. 

However, adhering to these rules often requires additional tokens in the prompts. For instance, to minimize error in output format, you might need to include a few-shot learning examples in the prompt to help the FM understand the format. These expanded prompts ensure that the communication format is well-defined, but they also increase the number of tokens used (token-overhead), which can be costly in terms of processing time and resources. 

To mitigate these costs, some solutions integrate classical robust-parsing techniques on the communication channels. Instead of spending too many tokens to ensure the quality of the communication protocol, these techniques can help parse FM responses more efficiently. A practical example of this is documented by Bottaro and Ramgopal~\cite{bottaro2024}, where they used a classic, CPU-powered robust YAML parser to detect errors in communication. This method helps maintain a low error rate (0.01\%), while also saving GPU jobs for more intensive tasks. By offloading the parsing to a CPU, they reduce the need for additional tokens in the prompt, leading to more efficient processing. Another example from Strong~\cite{instill2024} proposes a multi step pipeline approach to mitigate the correctness of the output structure where the output structuring step is separated out from actual model reasoning step to produce the correct structured output finally. But the proposed approach uses two inference calls which would increase both cost as well as latency.     

Offloading the output parsing and structure formatting to less costly CPU based solutions is a first step towards addressing this challenge. For instance, the output structuring step from Strong's work~\cite{instill2024} can be offloaded to a CPU before sending the result downstream. On the other hand, innovative decoding approaches (such as the one proposed by Beurer-Kellne et al.~\cite{beurer2024guiding}) which minimizes the performance overhead introduced with output structured generation, is another direction.

\noindent\textbf{Optimizing communication messages:} Recent approaches have identified ways to optimize communication by skip generating parts of the message that are already known. This allows one to avoid generating each token individually, especially when the structure of the response is predictable. 

For example, suppose we know that a response should have a format \texttt{<NAME="XXX">}. For a query like ``what is the name of the Nobel prize winner for peace in 2023'', the FM generates ``\texttt{<NAME=Narges Mohammadi>}." Instead of asking the FM to generate the entire response, we only ask the FM to generate the variable part (Narges Mohammadi). By using this approach, we can reduce the number of tokens that an FM needs to generate, leading to faster response times. 

A practical implementation of this concept is seen in the work of dottxt team~\cite{willkurt2024}. They proposed the Coalescence framework to speed up the inference by five times with their structured generation that skips unnecessarily calls to FMs leveraging the known structure of the responses, only generating the variable parts that change. This work proposes an efficient guided text generation technique using finite-state machines and regular expressions to enforce structural constraints, significantly reducing computational costs and enhancing output quality while being model-agnostic.

\subsection{\textbf{Challenge 3: The complexity of performance tuning and optimization of FMware}}
\label{challenge:FM_tuning}

\noindent Performance tuning and optimization in FMware requires a deep understanding of performance bottlenecks. The core of FMware is the inference of FMs. While many techniques focus on optimizing models~\cite{zhou2024survey}, efficiently serving FMs is only the beginning.
FMware involves interactions among multiple FMs and software components within a cognitive architecture, similar to classical software architecture, where each component has distinct resource demands.
This leads to numerous configuration knobs, further complicated by heterogeneous hardware. Additionally, FMware might evolve continuously by itself as its agents perform self-exploration, compared to regular software which is static. Optimizing live FMware is akin to hitting a moving target. Hence, we categorize the challenges into three dimensions as described below:

\noindent\textbf{Complex model-level optimization:} Techniques in FMware focus on enhancing hardware utilization, reducing latency, and maximizing throughput during the inference process. Existing FMs mostly rely on decoder-only transformer-based architectures. The inference process for these models has been described in detail in Section~\ref{sec:background}. Many optimization techniques have been proposed, including model architecture redesigns (e.g., multi-query attention) and model compression strategies (e.g., knowledge distillation, quantization)~\cite{wang2024model}. For a more in-depth understanding, the reader can refer to existing surveys~\cite{wang2024model,zhou2024survey}. In this section, we focus on the techniques that directly impact developers of FMware, where they interact with models through prompting.

In FMware, developers invest significant effort in crafting effective prompts for the FM, also known as prompt engineering. %\charles{old B1: Prompt Decomposition, Remove later}
Techniques like breaking down a complex prompt into multiple simpler prompts and adding explainability instructions can enhance model output quality and reliability. However, they may increase the number of model inference calls or output tokens, raising end-to-end latency. When chaining multiple FM innovations, prior tokens cannot be used by downstream FMs, causing waiting times between calls. In production, developers need to carefully balance these prompting techniques with their impact on overall performance.

Currently, prompt tuning relies heavily on manual and empirical methods. Developers frequently engage in trial-and-error approaches to refine prompts and find the optimal parameters. For instance, Chen et al.~\cite{chen2024design} reasoned about the pros and cons of task decomposition for LLM-based applications, where each task formats a prompt based on its input and feeds it into an LLM. They studied parallel decomposition to guide developers in achieving the expected accuracy or efficiency. To boost model inference performance, Kurt~\cite{willkurt2024} leveraged finite state machines and regular expressions to represent deterministic structures in the output, allowing the model to skip over predictable parts of the structure, thus substantially reducing generation latency. 
Streaming techniques have also been proposed to enhance performance by overlapping the output generation and input for the next model. For example, Bottaro and Ramgopal~\cite{bottaro2024} proposed streaming the application pipeline so that downstream calls can be invoked as soon as they are ready, without waiting for the complete response. Additionally, Santhanam et al.~\cite{santhanam2024alto} introduced ALTO, an FM serving system for streaming AI pipelines, demonstrating improved throughput and tail latency by streaming intermediate outputs to downstream tasks. 

While these methods are effective, the process is still manual and hard to extend to multiple objectives, making it hard to scale for more complex FMware. Future research might explore automating the prompt optimization process to minimize manual efforts. Through searching for multiple prompting goals such as output quality as well as performance requirements, the automated process enables developers to test and refine prompts rapidly. Additionally, real-world data analysis through matching the pairs of prompt templates and the outputs, can help with effective prompt designs for developers, providing fast turnaround times during prototyping. % Finally, creating prompting tools that consider both the quality of outputs and the performance of FM inference prompt generation will allow developers to focus on improving functional performance and facilitate more efficient prototyping.

\noindent\textbf{Excessive amount of performance configuration knobs:}
When dealing with FMware such as those illustrated by OPEA~\cite{opea_image} in Figure \ref{fig:rag_flow}, the configuration landscape becomes significantly complex. As shown, a performance engineer must consider multiple optimization opportunities, including different cognitive architectures, prompt designs, base model selections, model quantization decisions, fine-tuning processes, and communication protocol adjustments. These various aspects, from data ingestion to LLM inference and retrieval, illustrate the intricate array of configuration knobs required for optimizing the application-level performance. We describe three most prominent aspects as follows:

\begin{figure*}[htbp]
    \centering
    \includegraphics[width=\linewidth]{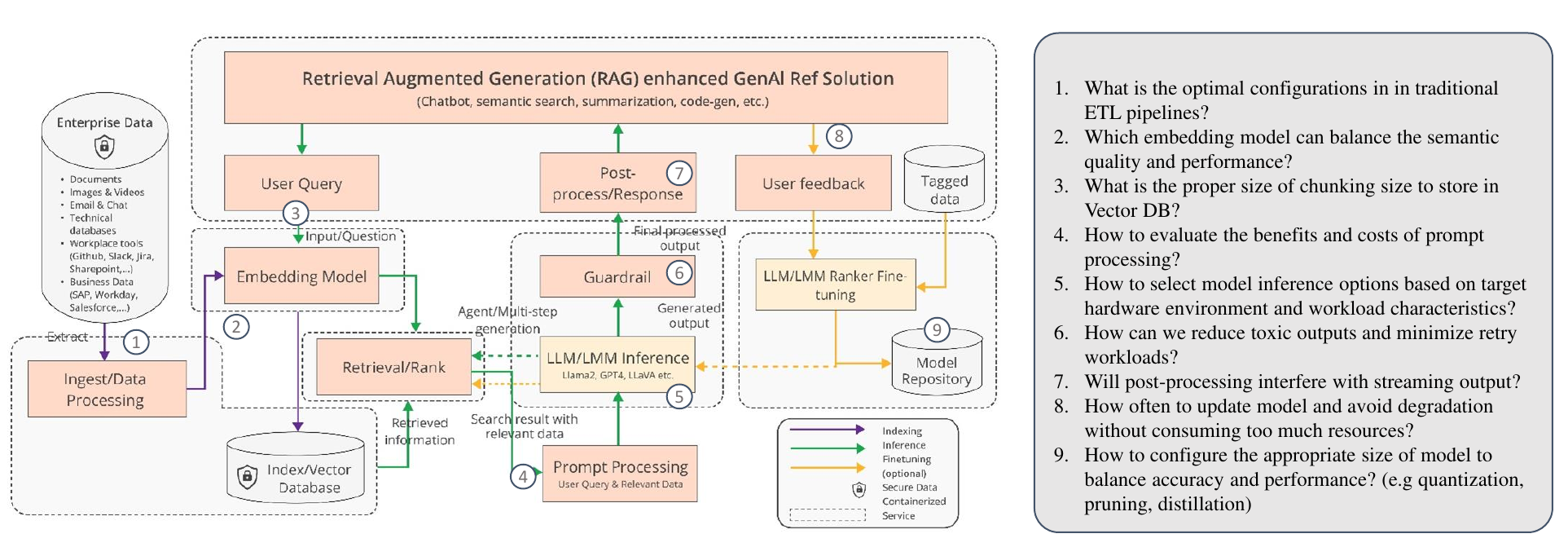}
\caption{A flowchart of the RAG-based LLM pipeline from OPEA~\cite{opea_image}, with added annotations on the right panel to highlight key tuning parameters and decision points, illustrating the complexities involved in FMware development}
    \label{fig:rag_flow}
\end{figure*}

Firstly, model selection involves a wide range of options, each performing differently in both functional and non-functional aspects such as generation speed, memory usage, and quality. Developers must not only aim for a model that produces good output quality but also one that satisfies pre-defined SLAs.

Secondly, the choice of inference engine must align with the hardware setup to either maximize throughput by leveraging hardware capabilities or minimize costs with CPU-based alternatives. Developers should recognize that different inference engines perform optimally under specific conditions.

Finally, the complexity increases when taking a holistic view of FMware's entire software stack. 
Optimizations must account for the costs of loading and unloading large FMs due to limited accelerator availability and high operational costs. When different teams work on separate parts of the same FMware and use different models, careful orchestration is required. Decisions such as workload splitting between CPU and GPU resources, selecting an appropriate model with an appropriate inference engine, and finding an optimal combination of these elements significantly affect system performance. For example, splitting workloads involves deciding which tasks are better suited for FM agents versus traditional software. However, the impact of these choices is not well-studied and rigorous engineering guidelines are lacking.

Many solutions have been proposed to address these challenges, each tackling a specific aspect of FMware optimization. Maurya et al.~\cite{maurya2024selectllm} proposed SelectLLM, a framework that analyzes user prompts and selects the most appropriate models at runtime. This approach enables developers to maintain response quality while reducing computational costs, thereby improving the efficiency of model selection. Similarly, Shekhar et al.~\cite{shekhar2024towards} introduced QC-Opt, a Quality-aware Cost Optimized LLM routing engine and framework. QC-Opt optimizes both the choice of LLM and input token count at runtime to minimize costs while maintaining output quality. This helps developers navigate trade-offs between quality and cost, providing flexibility in selecting models that best fit their requirements. Gong et al.~\cite{gong2024llm} developed a benchmarking toolkit to evaluate various quantization strategies and parameter configurations, providing insights that can guide developers in making decisions on pruning and optimizing models across different deployment scenarios.

To find suitable acceleration inference engines, while some articles provide high-level discussions and benchmarks for different engines ~\cite{bentoml_benchmarking}, there is still a lack of clear guidelines and standards to make informed selections. Further research and benchmarking are needed for specific scenarios. For instance, Xiao et al.~\cite{xiao2024large} investigated the pros and cons of MLC-LLM and Llama.cpp in mobile environments, using mobile-sensitive metrics such as battery power consumption, latency, and memory bottleneck. These insights are essential for tailoring LLM deployments to mobile devices where resource constraints are more stringent.

To track the complexities of FMware tuning in a holistic perspective, Sun et al.~\cite{sun2024cebench} provided a multi-objective benchmarking toolkit, CEBench, that focuses on balancing expenditure and effectiveness for LLM deployments. By allowing easy modifications through configuration files, CEBench supports holistic decision-making across the entire software stack, enabling developers to optimize resource allocation, cost, and performance in an integrated manner. Papaioannou et al.~\cite{papaioannou2024importance} proposed a holistic approach to tuning LLM applications by addressing the complexities introduced by diverse workloads and real-world conditions. They noticed that most LLM applications, which often rely on synthetic datasets, may not account for the variability in input sizes and task demands found in practical applications. Their analysis, which includes different workload types and memory configurations, helps identify key performance bottlenecks and optimization opportunities. By providing a framework that considers a wide range of use cases, their research guides developers in making more informed decisions to enhance FMware efficiency across various scenarios.

Addressing the complexity of performance optimization in FMware requires systematic studies, tools, and guidelines to support informed decision-making. Developing best practices, patterns, and anti-patterns can help developers determine which workloads are best suited to FMs versus traditional software approaches. Additionally, creating benchmarking tools or simulation platforms for comparing different models, parameters, and deployment environments would allow developers to test configurations quickly and assess their impact on performance and cost.
In the future, automated techniques, such as search-based multi-objective optimization, could further enhance productivity by autonomously tuning configurations to balance accuracy, latency, and cost. This approach reduces the need for time-consuming manual tuning while efficiently identifying optimal configurations, ultimately improving FMware performance.

\noindent\textbf{Evolving and moving target: }Unlike traditional software, FMware is live software that keeps evolving. Each round of execution of an agent might lead to an adjustment of the whole system. To make things worse, agents can self-evolve, making benchmarking much harder than traditional software. As a result, performance issues might be difficult to reproduce due to: (a) the probabilistic nature of the model inference process with token sampling, and (b) the evolving nature of FMware through Data Flywheel~\cite{oreilly2024llmtactical} or self-exploration.

To address the challenges of reproducibility, a common trick, also suggested by OpenAI~\cite{openai_managing_tokens}, is using settings like a low temperature parameter value to ensure more consistent inference outputs. Additionally, setting seeds beforehand, as suggested by PyTorch's reproducibility guidelines~\cite{pytorch_randomness}, can further help achieve consistent behaviour across repeated inferences. These methods help standardize model behaviour, simplifying the identification of performance bottlenecks in FMware. However, such reproducibility measures restrict the model's ability to autonomously explore and optimize.

The optimization process of FM and FM-powered agents should evolve from manually-tuning to a Data Flywheel-driven continuous self improving system. Firstly, we must continuously monitor and test if the accuracy of FMware drops through online feedback.
Machmouchi and Gupta~\cite{machmouchi_gupta_2023} proposed a comprehensive framework for evaluating LLMs, emphasizing the need for continuous testing with real-time user feedback. They highlighted the importance of segmenting user data to better capture the output quality of FMware. Such a framework would help us understand whether FMs need to be evolved.

The evolution of FMware requires developers to make decisions about how frequently to update the models. One possible solution is fine-tuning the models based on newly generated datasets. Alternatively, developers can update prompting and post-processing techniques to enhance output quality and save costs by not retraining the model. Developers must weigh the cost of fine-tuning as a significant investment against the use of efficient prompting and post-processing techniques applied at the individual request level. Xia et al.~\cite{xia2024understanding} proposed a profiling tool to help developers estimate the cost of LLM fine-tuning on GPU clusters, which aids in planning the frequency of fine-tuning based on cost considerations. The decision between these strategies often hinges on factors like user request volume and the desired performance level. In low-volume scenarios, investing in advanced prompting techniques can be more cost-effective because each request can afford additional computational time. Conversely, in high request volume environments, model fine-tuning becomes more beneficial as it allows simpler prompts, thereby reducing the computational overhead for each request.

For controlling the self-exploration behaviour, one approach is to first record the self-exploration and replay to reproduce the performance issue. Chen et al.~\cite{chen2022towards} proposed to reproduce the model training process with a record and replay mechanism. Similar ideas can be applied to FMware inference, e.g., the decoding process for executing each FM invocation
and other traditional software executions can be recorded and replayed as an agent explores, facilitating the analysis and debugging of performance issues in a reproducible manner.

\subsection{\textbf{Challenge 4: The complexity of deploying FMware}}
\label{challenge:deployment}

\noindent Several key decisions must be carefully considered when deploying FMware into production, as it must meet service-level agreements (SLAs). While many studies focus on ensuring the SLAs at the model level~\cite{oh2024exegpt, wu2023fast, gujarati2020serving}
% hui2024esg, romero2021llama, fu2024serverlessllm
, these efforts alone are insufficient to guarantee the FMware meets \textit{application-level} SLAs, since models are only part of the entire software system.
Driven by such requirements, we identify challenges in three dimensions described as follows:

\noindent\textbf{Selecting optimal deployment options when hosting FMware:} 
Unlike deploying traditional software, FMware deployment requires higher computation costs due to the invocation of FMs, which often involve the usage of specialized accelerators.
There are three types of deployment options for FMware: API-based deployment, rented cloud instances, and on-premise self-hosting. In certain cases, these deployment options can also be jointly leveraged.
API-based deployment follows a pay-as-you-go mechanism. Examples are OpenAI-compatible APIs by proprietary model providers~\cite{gemini} 
% zhipu_bigmodel
or Anyscale Model Endpoints API~\cite{anyscale_endpoints}. Developers send HTTP requests to served models to retrieve the generated tokens. While it is the simplest way to set up, the performance of API-based deployments solely relies on the API provider and can sometimes be unpredictable or unreliable~\cite{AIapocalypse2024}.
Rented cloud instances refer to renting computation resources from cloud service providers. Developers could either rent compute instances for dedicated purposes (e.g., AWS EC2 G5~\cite{aws_g5}) or in a serverless way~\cite{modal}.
% cloudflare_workers_ai
This option provides flexibility and can absorb spikes in request volume, as these platforms usually provide autoscaling mechanisms. However, it requires DevOps engineers to configure based on the computation requests. A common challenge is low hardware utilization resulting in unnecessary costs.
On-premise hosting means that a person or an organization procures physical or managed private cloud clusters which are dedicated to them. Such an option provides the maximum flexibility and control over hardware. At the same time, extensive engineering efforts are needed to guarantee the optimal usage of these hardware to satisfy the needs for multi-tenancy, as the resources usually need to be shared to cover the costs.
In terms of cost, API-based deployment can incur generation costs of roughly \$35k per month or \$420k per year, making it the largest single cost component for organizations~\cite{Roux2023_TotalCostOpenAI}. Cloud-based LLM deployments provide flexible pay-as-you-go pricing, but costs can become unpredictable at high usage levels and may reach 2-3 times the cost of on-premise solutions for large-scale operations~\cite{Latitude2025_CloudVsOnPremLLMCost}. On-premise self-hosting involves significant upfront investment but delivers more predictable costs and can yield 30-50\% savings over three years when hardware utilization consistently exceeds 60-70\%~\cite{Latitude2025_CloudVsOnPremLLMCost}.
The three above-mentioned deployment options have pros and cons. Hence, developers need to balance the degree of control over hardware, the costs, the utilization, and the expected application performance.

For API-based deployment, existing practices attempted mixed use of small and large FMs for latency reduction. % \boyuan{what is this build and minions? their system name? if so just mention xx proposed xxx, this expression is weird and maybe explain a bit more below?}
Both BiLD~\cite{kim2024speculative} and Minions~\cite{wang2024minions} showed that smaller models are effective in latency reduction (dropping to 50\% in the case of BiLD) with little-to-no output generation quality compromise. To control the unpredictability of APIs, Wang et al.~\cite{wang2024towards} studied the request and response token length distributions of ChatGPT and GPT4 models at the API level, and proposed that this trace can be used for optimizing serving systems to become ``workload-aware".

Rented cloud and on-premise hosting require a balance of performance and cost. Griggs et al.~\cite{griggs2024melangecostefficientlarge} remarked that the optimal GPU for cost-efficiency in running an FM varies and largely depends on the size of requests being processed. They showed that up to 77\% cost reduction is possible with heterogeneous GPU type selection at the time of deployment. 
Several industrial solutions like run:AI~\cite{runai_gpu_memory_swap} and Apache YuniKorn~\cite{apache_yunikorn_core} improve the utilization of accelerators through advanced scheduling features. However, none of the current solutions consider application-level SLA requirements.
Future work might explore the possibility of hosting FMware in a hybrid way. In combination with application-level SLA-aware scheduling~\cite{tan2024teola} and scaling algorithms~\cite{griggs2024melangecostefficientlarge, rayservepow2}
% \boyuan{which work talks about scaling? if not just cite ray serve}
, more research is needed to achieve the optimal performance and lowest costs for deploying FMware.

\noindent\textbf{Deploying multi-process FMware efficiently:}
FMware comprises multiple concurrent processes on a unified cluster (uni-cluster) to enhance performance: inference for serving models, data flywheel for fine-tuning and updating models, and agent self-exploration for autonomous planning. These processes share computation and bandwidth resources. 

Scheduling these processes efficiently involves selecting compatible ones for co-location, which requires understanding the characteristics of each process to avoid cross-process interference. Processes may also be described as \textit{inertial}, meaning once started it is costly to preempt or revert the state, due to the scale of data being manipulated or transferred. Model weights loading before inference is one such example while the data-loading phase of training/fine-tuning is another; both require a significant bandwidth of the PCIe system bus. Therefore, co-locating these two processes could cause performance issues. To mitigate interference, one can time-slice processes with preemption as traditional schedulers do. 

Even within one process such as inference, space separation at this finer granularity is shown to be effective. Disaggregation was applied by Hu et al.~\cite{hu2024inference} to separate prefill and decode instances, improving the performance over cost metric by 2.4x. Memory capacity is also a scarce resource in this scenario, and the variability of prompt and output tokens for one inference process leads to variability in leftover memory for other processes. This is largely due to the KV cache memory used for each output token, further complicating memory allocation and scheduling strategies to attain maximum throughput in serving systems. 
Cheng et al.~\cite{cheng2024enabling} proposed a ``Wasted Memory Access" (WMA) metric to accurately predict memory consumption, so that corresponding memory required at certain batch sizes of inference requests can be leveraged for smart scheduling decisions.

When dealing with complex cognitive architectures, these issues are further amplified. Unlike traditional runtime systems where resources need to be predefined, Agentware requires dynamic allocation, as agents operate autonomously without defined code paths and share runtime resources on the fly. Under a resource-constrained environment, this can lead to contention and interference among agents, impacting other processes' performance. New ``OS-like" architectures have been proposed, where FMs act as the kernel to govern access to shared hardware resources and services~\cite{mei2024llm,li2024personal}, to make agent completion latency predictable. In addition,
Mei et al.~\cite{mei2024llm} show that isolation between \textit{modules} in the \textit{LLM Kernel} is the key to preventing resource conflicts with the rest of the system, and ensuring optimal access to resources and services when agents execute tasks.

The existing ``Model-as-a-Service" paradigm is inadequate to capture the multi-process nature of FMware. Aggressive queue-based approaches risk over-provisioning, while SLA-aware scheduling and resource provisioning algorithms on the application level can better match resources to latency targets and execution trends. 
Future work should focus on optimizing time-slicing and spatial disaggregation for FMware processes, determining the ideal separation granularity for both inter-process (e.g., training/inference) and intra-process (e.g., prefill/decode in inference) operations.

\noindent\textbf{Deploying multi-tenant FMware efficiently:} 
Current infrastructures and hardware accelerators are too expensive to dedicate to a single FMware, necessitating a need for multi-tenant optimization objectives. %Single-tenant objectives focus on maximizing the performance and cost-effectiveness of FMware for individual users. However, 
It is often economical to have a cluster shared among multiple deployments, which introduces the challenge of multi-tenancy. The primary objective is to maximize the cluster-level hardware utilization and efficiency of the shared hardware across all FMware deployments while trying to meet every tenant's performance goals; each tenant may have a different volume of users and SLA requirements. In some scenarios, there can be conflicting performance requirements when diverse types of FMware are co-located, as such a uniform cluster-wide policy is not optimal. Lazuka et al.~\cite{lazuka2024llm} illustrated the difficulty of simultaneously meeting a low-latency SLA required by chatbot workloads and a high throughput SLA required for text summarization workloads, within the same serving system.

One way to share the cluster is through sharing served FMs across different FMware and keeping models persisted in memory to avoid model loading costs, as well as reducing the occupied memory. Selecting batch sizes of requests to optimize for both latency and throughput across many deployed FMs becomes challenging. This is particularly important for Promptware, where multiple FMs may be used in a pipeline or workflow of invocations, under end-to-end SLA constraints. %This necessitates strategic decisions about when to create or delete model replicas to optimize overall performance. 
Tan et al.~\cite{tan2024teola} showed that topology-aware batching at the application level can achieve a latency reduction of up to 19\% under multi-query workload scenarios. The batching is performed based on workflow dependency of multiple requests to meet multiple SLAs simultaneously, and a batch size is selected to be most efficiently processed by the execution engine. However, scaling strategies when existing resources are not enough to meet request SLAs were not considered in the scope of this work.

In addition to batching, optimally routing requests to compute resources is another critical aspect.
Lin et al.~\cite{lin2024parrot} demonstrated that routing requests with a common prefix to the same inference instance maximizes the reuse of existing KV cache, thus achieving high locality and utilization. Their proposed \textit{Semantic Variable} is a declarative approach at the FMware level for schedulers to be optimized around user intent. These strategies also help improve memory usage predictability and overall performance stability, since the size of KV cache used is known during the routing process, instead of allocating the cache anew. To minimize data movement, Sun et al.~\cite{sun2024llumnix} proposed a live migration mechanism to enable runtime rescheduling, resulting in 26x lower latency at the prefill phase. This is accomplished using a two-level global and instance-level scheduler. The former coordinates request dispatching, migrations, and autoscaling actions, and the latter reports the memory load and virtual usage of each instance back to the global scheduler. Both works are limited however as they do not simultaneously consider SLA aspects for scaling decisions at the FMware level, only at each inference request from a ``Model-as-a-Service" understanding.

Existing studies remain siloed, lacking integration among model-level, multi-tenant optimization and cluster elasticity. Future research should explore advanced methods for understanding user behaviour, intent~\cite{lin2024parrot}, and interference sensitivity. Expanding beyond ``cluster-in-a-vacuum"-scale to internet-scale awareness~\cite{AIapocalypse2024} would enable intelligent, flexible policies to anticipate performance degradation for each tenant. FMware serving should jointly consider scheduling (assigning requests to model replicas) and resource allocation (determining the number of replicas to deploy) to meet multiple performance objectives at the application level.  In the next section, we discuss our system design to achieve these goals.

\endgroup

\section{Our vision towards an SLA-aware \fwo}
\label{sec:runtime}
\subsection{\fwo Design}
\label{sec:runtime:architecture}

In this section, we present the design, architecture, and implementation details of our performance-oriented runtime (\fwo), which is built to serve multi-tenant FMware. Our approach prioritizes SLAs as the central design principle, aiming to satisfy each FMware's SLA requirements while optimizing cluster-level hardware utilization.

Our design is specifically tailored for Promptware, not Agentware. Agentware’s dynamic, autonomous control flow and evolving behavior introduce complexity (such as non-deterministic execution and recursive self-correction) that falls outside the scope of this study. Instead, our focus is on Promptware, where the application is represented as DAGs. The nodes of the graph represent tasks and the edge represents the control flow dependencies (e.g., sequential or conditional branching). To simplify the description, we treat each task as an invocation to an FM. However, in practice, the tasks can also take other forms, such as regular code execution or external API calls served outside our runtime.

Our runtime can be deployed on an on-premise cluster or a rented cloud cluster, provided we have full access to the machines. While individual machines are equipped with the same accelerator setup (e.g., 8 Ascend 910B NPUs~\cite{liang2020ascend} per machine), the runtime is fully aware of and leverage the cluster-wide heterogeneity. For each FMware workflow deployed on \fwo, the workflow is first represented as a DAG. During execution, this DAG is decomposed into concrete model requests, meaning each node that involves an FM invocation is translated into one or more calls to the relevant model(s). For example, a workflow request could be represented as a sequential DAG with nodes invoking Model A → Model B → Model C, where the outputs of one node can serve as inputs to downstream nodes. Models can also be reused across multiple workflows, as discussed in Section \ref{challenge:deployment}, allowing shared FM resources to serve different workflow requests efficiently.

\begin{figure}[htbp]
    \centering
    \includegraphics[width=0.5\linewidth]{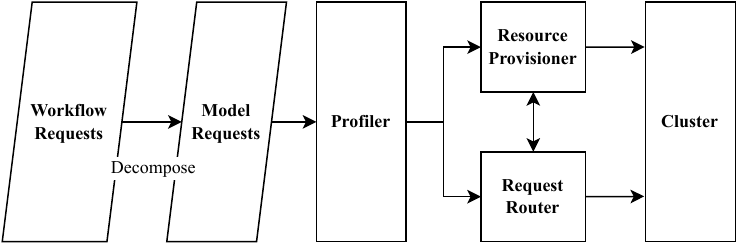}
    \caption{The simplified architecture and components of \fwo}
    \label{fig:FWO}
\end{figure}

The simplified architecture and the core components of \fwo are illustrated in Figure~\ref{fig:FWO}. To meet performance requirements, we present the four core components of our system architecture:

\begin{itemize}%[leftmargin=*,itemindent=1em,align=left,noitemsep,topsep=0pt]
    \item \textbf{Profiler.} This component is responsible for offline profiling of each FM on each available hardware type. It calculates the estimated inference latency (given a default number of tokens to ensure determinism) as well as memory consumption for each type of FM invocation. This profile data is crucial, as it provides referenced measurements used by the runtime. For example, the referenced latency can be used as an estimation of how much time each task should consume so that it will not impact the end-to-end latency goal. It also translates the end-to-end workflow SLO into per-task time budgets, which we refer to as ``slack'' -- a portion of the total SLO allocated to each task. Profiling is done offline, and each FM is profiled only once.
    \item \textbf{Resource Provisioner.} This component acts as the coarse-grained control for SLO adherence. It makes strategic decisions about allocating or releasing accelerator resources based on the monitored risk of SLA violations. If workflows are frequently exhausting their slack and risk missing end-to-end SLOs, the Resource Provisioner checks for available cluster resources to spin up new FM replicas. Conversely, if a replica remains idle for a configured period, it will be terminated to free up resources.
    \item \textbf{Replica Router.} This component is the fine-grained control and provides the core decision-making function of \fwo. It is responsible for routing each individual model request to a specific model replica. As the Resource Provisioner creates more replicas (potentially on different hardware types), the Replica Router must intelligently decide which replica should serve an incoming request.
    To achieve this, the Replica Router implements a novel, SLO-aware scheduling algorithm (Algorithm~\ref{alg:FWOSchdule}). Unlike static routing (e.g., round-robin), our scheduling mechanism dynamically considers user SLOs (via slack) and the heterogeneous card types available.
    The algorithm's core is an objective latency cost function that decides the optimal hardware configuration (i.e., which specific replica on which card type) to use for each node invocation in the workflow. This cost function estimates the total latency for a request on a specific hardware type by combining three key factors: (a) Estimated Inference Latency $lx$: The base execution time for the model on that specific hardware, derived from the Profiler; (b) Queuing Delay $lq$: The estimated time the request would spend waiting in the target replica's queue; (c) In-flight Request Delay $lr$: The processing time remaining for requests already being executed (in-flight) by that replica.
    The final estimated latency $l_{est}$ is the sum of these factors: $l_{est} = lx + lq + lr$. For every incoming request, the Replica Router iterates through all possible configurations (i.e., all available replicas across all hardware types), calculates $l_{est}$, and selects the configuration that minimizes this estimated latency. This ensures that each request is routed to the replica most likely to meet its slack requirement, thereby protecting the end-to-end workflow SLO.
    \item \textbf{Cluster.} This component is the execution layer that handles commands from the Resource Provisioner and Replica Router. It manages cross-node communication, data movement, and the lifecycle of model replicas (e.g., loading model weights into accelerator memory).
\end{itemize}

\begin{algorithm}
\caption{Scheduling Algorithm}
\begin{algorithmic}[1]
\State {$n$: node to be scheduled}
\State {$C$: list of profiled configurations for the node}
\State {$slack$: estimated slack for the node}
\Procedure{FWOSchedule}{$n, C$, $s$}
    \State $costs\_target$ = \{ \}, $costs\_greedy$ = \{ \}
    \For{$c1$ in $C$}
        \State $lx$ = ProfiledLatency($c1$)
        \State $card$ = $c1.card\_type$
        \State $lq$ = QueingDelay($card$)
        \State $lr$ = InflightRequestsDelay($card$)
        \State $cost$ = $lx + lq + lr$
        \If{$cost < slack$}
            \State $costs\_target.put(cost)$
        \Else
            \State $costs\_greedy.put(cost)$
        \EndIf
    \EndFor
    \State $config$ = Min($costs\_target, costs\_greedy$)
    \State $n.set\_config(config$)
\EndProcedure
\end{algorithmic}
\label{alg:FWOSchdule}
\end{algorithm}

The design of \fwo prioritizes robust end-to-end SLO enforcement, leading to several key trade-offs. First, we applied a dynamic, SLO-aware scheduling algorithm over simpler static load balancers (like Ray Serve's baseline). While static approaches are simpler to implement, they are SLO-agnostic and poorly handle hardware heterogeneity, leading to SLO violations. We traded this simplicity for a higher scheduling complexity in our Replica Router, which is essential for fine-grained control. Secondly, our Replica Router operates within a logically centralized control plane to maintain a global view of all workflows. The trade-off is that a decentralized design might offer greater scheduler scalability, but it sacrifices the global context necessary to make optimal decisions for end-to-end SLOs. Finally, within our Replica Router, we chose a cost function considering three aspects ($l_{est} = lx + lq + lr$) instead of simpler heuristics such as the shortest queue. While computationally cheaper, heuristics ignore crucial factors like hardware differences (e.g., a short queue on a slow card) and in-flight request state. We traded this lower computational cost for the superior accuracy of our approach, which provides a holistic latency estimate critical for making correct scheduling decisions in a heterogeneous environment.

\textbf{Discussion.} The \fwo architecture is explicitly designed to solve the complexities of Challenge 3 (Performance Tuning and Optimization) and Challenge 4 (Deployment). The Profiler addresses Challenge 3 by converting the complex, multi-knob optimization problem into a singular slack metric, simplifying performance optimization. It also provides the foundational profiling data needed to address the evolving nature of performance targets. The Resource Provisioner tackles Challenge 4's multi-process multi-tenancy issue by providing dynamic elasticity, ensuring balanced resource usage by scaling resources based on risk of SLA violation. Finally, the Replica Router provides the core solution: its SLO-aware scheduling algorithm leverages the cost function to algorithmically optimize performance (Challenge 3) and make efficient, real-time routing decisions required for stable multi-tenant deployment (Challenge 4) across heterogeneous hardware.

\subsection{Quantitative Evaluation}
\label{sec:runtime:architecture}

% %%%
% HZ: this paragraph is replaced by the following lines
% %%%
% We have developed a prototype system based on the design described above, which has already been deployed internally in production in a cloud environment. Initial evaluation results demonstrate that our SLA-aware \fwo outperforms established open source solutions like Ray Serve~\cite{ray_serve}, which can also handle routing requests and scaling up instances when needed. However, without considering SLA constraints, existing solutions would experience a higher SLA violation rate as the request load increases. In the future, we plan to extend the system to cope with other aforementioned challenges.

To validate our proposed design of \fwo, we have developed and deployed a complete implementation of the FMware Runtime internally in a production cloud environment. We conduct a quantitative evaluation for the SLA-aware \fwo, comparing its performance against Ray Serve~\cite{ray_serve}, a widely-adopted open source solution for FM serving that can also handle routing requests and scaling up instances when needed. In a recent survey, the only multi-model serving framework specifically mentioned is Ray, which supports serving multiple LLMs, underscoring the limited set of directly comparable baselines~\cite{xu2024survey}.

\subsubsection{Evaluation Setup}

% \noindent\textbf{Experiment Setup.} We deploy both \fwo and Ray Serve on identical hardware configurations with 2 nodes, each including 8 accelerator cards (in total of 480GB accelerator memory). The test workload consists of two FMware applications that are deployed concurrently (with three and two models, perspectively), requesting to three unique FMs: a Baichuan model (model size: 7 GB), a Llama-2-7b-hf  model (model size: 7 GB), and a Qwen1.5-7B-Chat model (model size: 7 GB). We conduct load test under a varying load condition of incoming requests that follow the Poisson distribution controlled by the number of requests per second. The SLA target is set to 130 and 90 seconds for two applications: wf1 and wf2, respectively. We measure the goodput rate, that is, the proportion of requests that meet the SLA target.

The evaluation was conducted on a heterogeneous computing cluster to realistically assess \fwo's capability to manage hardware resources under SLA. The cluster comprised two nodes with distinct accelerator card types: one node featured a faster card type, and the other a slower card type.

We deployed two distinct FMApp workflows to comprehensively test the system:
\begin{enumerate}[label=\arabic*.]
    \item Retrieval-Augmented Generation (RAG) Application: This simple workflow involves an initial model performing information fetching and lightweight computation, followed by a second model executing a more complex, computation-heavy generation task.
    \item Industry-based FMApp: This represents a customer-facing shopping chatbot workflow. It is characterized by conditional and model nodes, executing different DAG paths based on user input, potentially invoking up to five model calls.
\end{enumerate}

For both workflows, the SLA target was estimated based on the Time To First Token (TTFT). The target was defined as the TTFT measured under optimal conditions (i.e., without hardware competition) plus an acceptable buffer time, setting a realistic user expectation for initial output delivery.

To ensure reproducible testing for controlled and deterministic workloads, we used a very low temperature setting and no additional parameters during the model's generation, mitigating variations from random token generation. Conversely, for testing under realistic and nondeterministic workloads, we utilized a more typical temperature and realistic model parameter settings to accurately simulate real-world user experiences. Load testing was performed by sending requests following a Poisson distribution, simulating natural, variable user arrival rates.

\subsubsection{Evaluation Result}

% \noindent\textbf{Experiment Result.} Our experiment result demonstrates significant improvements of \fwo's SLA-aware scheduling for SLA compliance. Figure~\ref{fig:goodput} shows that under unsaturated load (i.e., up to 50 requests per second), \fwo can sustain a nearly 100\% goodput, while Ray Serve has a decaying goodput (a significant portion of requests violate SLA). Even when the request rate is higher (up to 100 requests per second), \fwo can sustain a significant higher goodput rate compared with Ray Serve. The better performance of \fwo can be attributed to its proactive scaling based on assessment of SLA satisfaction and its requesting routing algorithm. Our evaluation results demonstrate that our SLA-aware \fwo outperforms Ray Serve, particularly in maintaining service quality under variable load conditions. In the future, we plan to extend \fwo to handle more complex scenarios (e.g., heterogeneous hardware) and cope with other aforementioned challenges.

% \begin{figure}[htbp]
%     \centering
%     \includegraphics[width=\linewidth]{fig/goodput.png}
%     \caption{The simplified architecture and components of \fwo}
%     \label{fig:goodput}
% \end{figure}

We evaluated the performance of \fwo against the Ray Serve baseline, focusing first on the latency aspect. The results demonstrate that \fwo's dynamic, profile-aware dynamic scheduling provides a significant reduction in latency under both deterministic and nondeterministic conditions. We evaluated two primary scheduling modes: a greedy mode, which prioritizes immediate resource optimization, and a target mode, which balances load to meet SLA targets.

\noindent\textbf{Latency.} Under controlled, deterministic workloads, \fwo achieves a substantial reduction in the request execution latency before the system reaches saturation. Compared to the baseline, the greedy and target modes reduce the average request execution latency by 33.81\% and 30.7\%, respectively. This improvement stems from \fwo's performance-oriented scheduling. We analyzed the scheduling decisions in a 15-request test case. The baseline system, Ray Serve, routed only 30\% of the model executions to the faster accelerator card. In contrast, \fwo's greedy mode assigned 70\% of executions to the faster card, while the target mode assigned 36.7\% to balance throughput. Even though Ray Serve provides features such as static fractional GPU and auto-scaling in a fixed range to ensure the cluster’s performance efficiency, the nondeterministic nature of FMApp execution makes it difficult to adopt a static scheduling mechanism. On the other hand, our profile-aware dynamic scheduling in \fwo: (1) leverages the profiled performance metadata of model execution on different accelerator card types, and (2) uses a dynamic scheduling algorithm together with a real-time feedback loop mechanism to efficiently execute FMApp requests.

We also measured TTFT, a critical metric for user-perceived latency. \fwo reduced the average TTFT by 33\% (greedy mode) and 30\% (target node) across workloads from 1 to 30 requests.

To simulate a more realistic environment, we introduced nondeterminism in model generation, where inference times varied even with an identical max new token setting. In this scenario, \fwo's robustness became evident. Table~\ref{table:realistic_compare} shows \fwo consistently outperformed the baseline, achieving a lower median TTFT across all tested request loads, from a single request up to system saturation. This result confirms that our profile-driven dynamic scheduling approach adapts effectively to the inherent variability of model inference in real-world conditions.

\begin{table}[h]
\centering
\caption{Compare the median TTFT and goodput rate across different numbers of requests among Ray Serve, \fwo in target and greedy modes.}
\label{table:realistic_compare}
\begin{tabular}{|l|c|c|}
\hline
System & Median TTFT & Median Goodput Rate \\ \hline
Ray Serve & 153.6 & 0.53 \\ \hline
\fwo (greedy) & 147.9 & 0.54 \\ \hline
\fwo (target) & 126.0 & 0.61 \\ \hline
\end{tabular}
\end{table}

\noindent\textbf{Goodput.} Besides latency reduction, \fwo demonstrates better efficiency in terms of goodput rate, defined as the proportion of requests successfully completed within the specified SLA target --- a metric to quantify effective request completion under SLA constraints~\cite{Wang2024_RevisitingSLOGoodput}. In the same experiment, we achieved a significantly higher goodput rate compared to the baseline, with an average increase of 26.39\% in the greedy mode and 15.27\% in the target mode when serving our studied FMApp case before system saturation. This enhanced performance translates directly to increased capacity for SLA-compliant service. For instance, before engaging any scaling strategies, using a 70\% goodput rate threshold as an acceptable service quality boundary, Ray Serve was capable of processing 7 requests. In contrast, \fwo could process 8 requests in the target mode and 10 requests in the greedy mode. This indicates that \fwo more efficiently utilizes hardware resources, achieving a better goodput rate through its dynamic resource management. Furthermore, Table~\ref{table:realistic_compare} shows that \fwo maintains its high goodput rate even in the aforementioned realistic, nondeterministic scenario.

\section{Conclusion}
\label{sec:conclusion}
This paper has explored the emerging SPE challenges for FMware, highlighting numerous opportunities for innovation to enhance current practices. We explored four major challenges spanning the software lifecycle for FMware and discussed our attempt to addressing SLA-aware multi-tenant serving.
Our vision of \fwo marks only the first step towards tackling these challenges, stemming from our experience working with FMware developers to resolve performance issues, discussions with world-renowned scholars, and comprehensive surveys. The insights presented here aim to help developers address performance concerns more effectively than traditional SPE methodologies.
Future work should focus on establishing benchmarks and tooling to enable researchers and practitioners to systematically observe real-world FMware performance issues, empirically validate these four SPE challenges, and optimize performance across a broader set of Promptware and Agentware projects.
We encourage both researchers and practitioners in the SPE community to advancing FMware performance engineering. The unique challenges outlined in this paper represent critical areas for future work, as the SPE field continues to evolve alongside next-generation AI-powered software.

\section{Disclaimer}
\label{sec:disclaimer}
Any opinions, findings, conclusions, or recommendations expressed in this material are those of the author(s) and do not reflect the views of Huawei. Also, ChatGPT was used for copy-editing. All experiments, analysis, writing, and results were performed by the authors, who also thoroughly reviewed the final content. This complies with IEEE and ACM policies on AI use in publications.

\balance
\clearpage

%%
%% The next two lines define the bibliography style to be used, and
%% the bibliography file.
\bibliographystyle{ACM-Reference-Format}
\bibliography{reference}

\end{document}